\documentclass[useAMS,usenatbib]{mn2e}

\usepackage{graphicx}
\usepackage{psfig}
\usepackage{natbib}
\usepackage{booktabs}
\usepackage{amsmath}
\usepackage{amssymb}
\usepackage{txfonts}
\title[A possible compact companion to HD 164816]{Is there a compact companion orbiting the late O-type binary star HD 164816?}
\author[L. Trepl et al.]{L. Trepl$^{1}$\thanks{E-mail: ludwig.trepl@uni-jena.de}, V.V. Hambaryan$^{1}$ , 
T. Pribulla$^{2}$\thanks{E-mail: pribulla@ta3.sk}, N. Tetzlaff$^{1}$, R. Chini$^{3,4}$, R. Neuh\"auser$^{1}$,\\ 
\LARGE{\rm{S. B. Popov$^{7}$, O. Stahl$^{5}$, F. M. Walter$^{6}$}, M. M. Hohle$^{1}$}\\
$^{1}$Astrophysikalisches Institut und Universit\"ats-Sternwarte, Universit\"at Jena,
Schillerg\"a\ss chen 2-3, 07745 Jena, Germany\\
$^{2}$Astronomical Institute, Slovak Academy of Sciences, 059~60
Tatransk\'a Lomnica, Slovakia\\
$^{3}$Astronomisches Institut, Ruhr-Universit\"at Bochum, Universit\"atsstr. 150,
44801 Bochum, Germany\\
$^{4}$Instituto de Astronom\'ia, Universidad Cat\'olica del Norte, Antofagasta, Chile\\
$^{5}$ZAH, Landessternwarte, K\"onigstuhl 12, D-69117 Heidelberg, Germany\\
$^{6}$Department of Physics and Astronomy, Stony Brook University, Stony Brook NY\\
$^{7}$Sternberg Astronomical Institute, Lomonosov Moscow State University, Universitetski pr. 13, Moscow, 119991, Russia}

\begin{document}

\newcommand\hd{HD~164816}
\newcommand\xmmssc{2XMM~J180356.8-241845}

\date{Accepted 2012 August 29. Received 2012 August 29; in original form 2011 October 06}

\pagerange{\pageref{firstpage}--\pageref{lastpage}} \pubyear{2002}

\maketitle

\label{firstpage}

\begin{abstract}
We present a multi-wavelength (X-ray, $\gamma$-ray, optical and radio) study of \hd, a late O-type X-ray 
detected spectroscopic binary. X-ray spectra are analyzed and the X-ray photon arrival times are checked for 
pulsation. In addition, newly obtained optical spectroscopic monitoring data on \hd~are presented. They are 
complemented by available radio data from several large scale surveys as well as the \emph{FERMI} $\gamma$-ray 
data from its \emph{Large Area Telescope}. We report the detection of a low energy excess in the X-ray spectrum 
that can be described by a simple absorbed blackbody model with a temperature of $\sim$ 50 eV as well as a 9.78 s 
pulsation of the X-ray source. 
The soft X-ray excess, the X-ray pulsation, and the 
kinematical age would all be consistent with a compact object like a neutron star as companion to HD 164816.
The size of the soft X-ray excess emitting area is consistent with a circular region with a radius of about 7 km,
typical for neutron stars, while the emission measure of the remaining harder emission is typical for late O-type 
single or binary stars. If HD 164816 includes a neutron star born in a supernova, this supernova
should have been very recent and should have given the system a kick, which is consistent with the observation
that the star HD 164816 has a significantly different radial velocity than the cluster mean. 
In addition we confirm the binarity of \hd~itself by obtaining an orbital period of 3.82 d, 
projected masses $m_1 {\rm sin}^{3} i$ = 2.355(69) M$_\odot$, $m_2 {\rm sin}^{3} i$ = 2.103(62) M$_\odot$ apparently 
seen  at low inclination angle, determined from high-resolution optical spectra.
\end{abstract}

\begin{keywords}
stars: neutron, stars: individual: \hd, stars: individual: \xmmssc.
\end{keywords}

\section{Introduction}

The expected total number of Neutron stars (NSs) in our Galaxy is predicted to be
10$^{\rm 8}$-10$^{\rm 9}$ (Narayan \& Ostriker 1990) of which isolated NS should form
the majority. Until today there are
$\sim$2000 known radio pulsars and only seven known isolated thermally emitting NS (INS)
called the Magnificent Seven (Haberl 2007, Kaplan et al. 2011). Since the discovery of the first INS (RX J1856.5-3754)
in 1996 (Walter et al. 1996) the search for more thermally emitting NSs is an ongoing
process. Those seven objects have been recognized by their high X-ray to optical flux
ratio and their rather soft X-ray emission represented by low X-ray hardness ratios. Thus
looking for objects with similar properties is one way to find new candidates, see e.g.
Pires et al. (2009). It is clear that many candidates can be missed, for example when they are harbored
in binary or multiple star systems, since the X-ray flux is dominated by the host star, or as compact companions to 
runaway stars (Posselt et al. 2008). Searches for Pulsar companions around OB runaway stars have been performed 
by Philp et al. 1996, Sayer et al. 1996.\\
We search for X-ray pulsations from all X-ray sources near or identified with galactic OB stars, in search
of non-interacting (and hence effectively
isolated) NSs that remain bound in a stellar system following the supernova. About 10\% of such
systems remain bound after the first supernova (Kuranov et al. 2009).\\
HD 164816 is located in the direction of the young open cluster
NGC 6530. Prisinzano et al. (2005) recently investigated in depth
the distance of this cluster: Literature values range from 560 pc
(Loktin \& Beshenov 2001) to 2000 pc (Walker 1957, van den Ancker et al. 1997).
The mean of the distance estimates (in table 1 of Prisinzano et al. (2005)
and the value found by Prisinzano et al. (2005) themselves, 1250 pc)
is $1543 \pm 345$ pc, which we now use in this paper; all but one
(560 to 711 pc, Loktin \& Beshenov 2001) values found in the
literature are consistent with this value within $1~\sigma$. 
Within $1~\sigma$, our value is also consistent with the distance  
of $\sim 1250$ pc obtained by Prisinzano et al. 2005.

HD 164816 itself is located at $\alpha$ = 18h 03m 56.866s
and $\delta = -24^{\circ} 18^{\prime} 45.22^{\prime \prime}$
and has V=7.09 mag (Reed 2003). Recently, Megier et al. (2009)
determined the distance towards HD 164816 by interstellar Ca II
absorption to be $864 \pm 136$ pc. Hence, the distances of the
NGC 6530 cluster and the star HD 164816 are deviant by 1 to $2~\sigma$,
which may indicate that HD 164816 is somewhat foreground to the cluster,
to be discussed below.\\ 
The region 
around it is also known as the Lagoon Nebula (NGC 6523 = M 8) which is one of the brightest HII regions 
in the Galaxy where star formation has started a few 10$^{7}$ yrs ago (van den Ancker 1997). However, 
the membership of \hd~is uncertain.\\
NGC 6530 has about 2000 known members which display an age gradient (Damiani, Prisinzano, Micela \& Sciortina
2006). From March 2001 to July 2003 \hd~has been in the field of view of seven X-ray
observations, one with \emph{XMM Newton} and six with the Chandra X-ray observatory
(\emph{CXC}).\\
In the following we will present the X-ray spectral and temporal properties from
\emph{XMM PN} and \emph{Chandra ACIS-I} data indicating a compact companion candidate (\S 2). We present 
the orbital parameters of the spectroscopic binary O-type star in \S 3. Then we present the 
available radio survey data as well as the analysis of $\gamma$-ray data obtained by \emph{FERMI's} 
Large Area Telescope (\emph{LAT}) in \S 4. A discussion of the physical nature of the possible companion will
be given in \S 5 and concluding remarks will be made in \S 6.

\section[]{X-ray data Analysis}
\begin{figure*}
  \centering
  \includegraphics[width=\columnwidth]{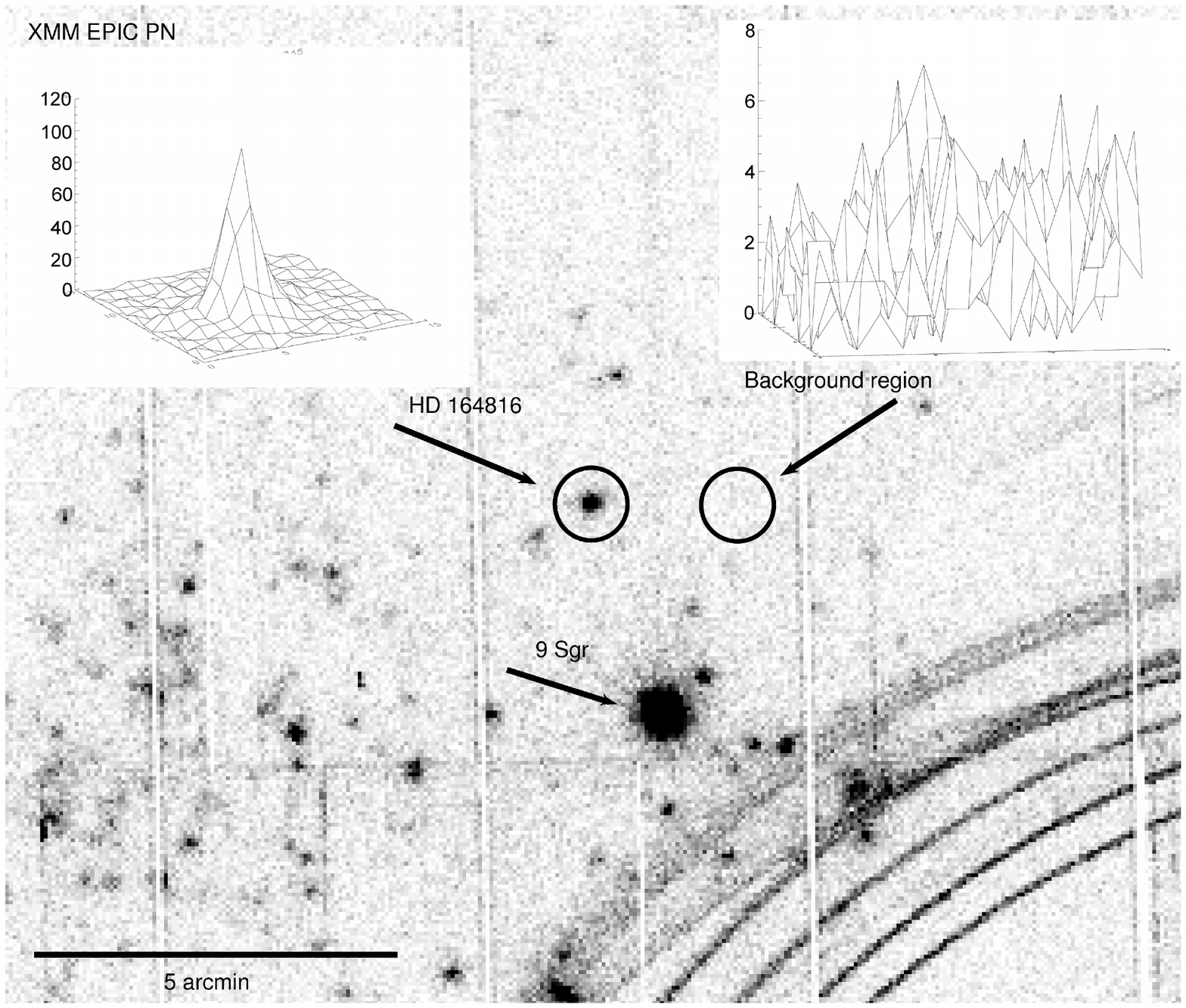}
  \includegraphics[width=\columnwidth]{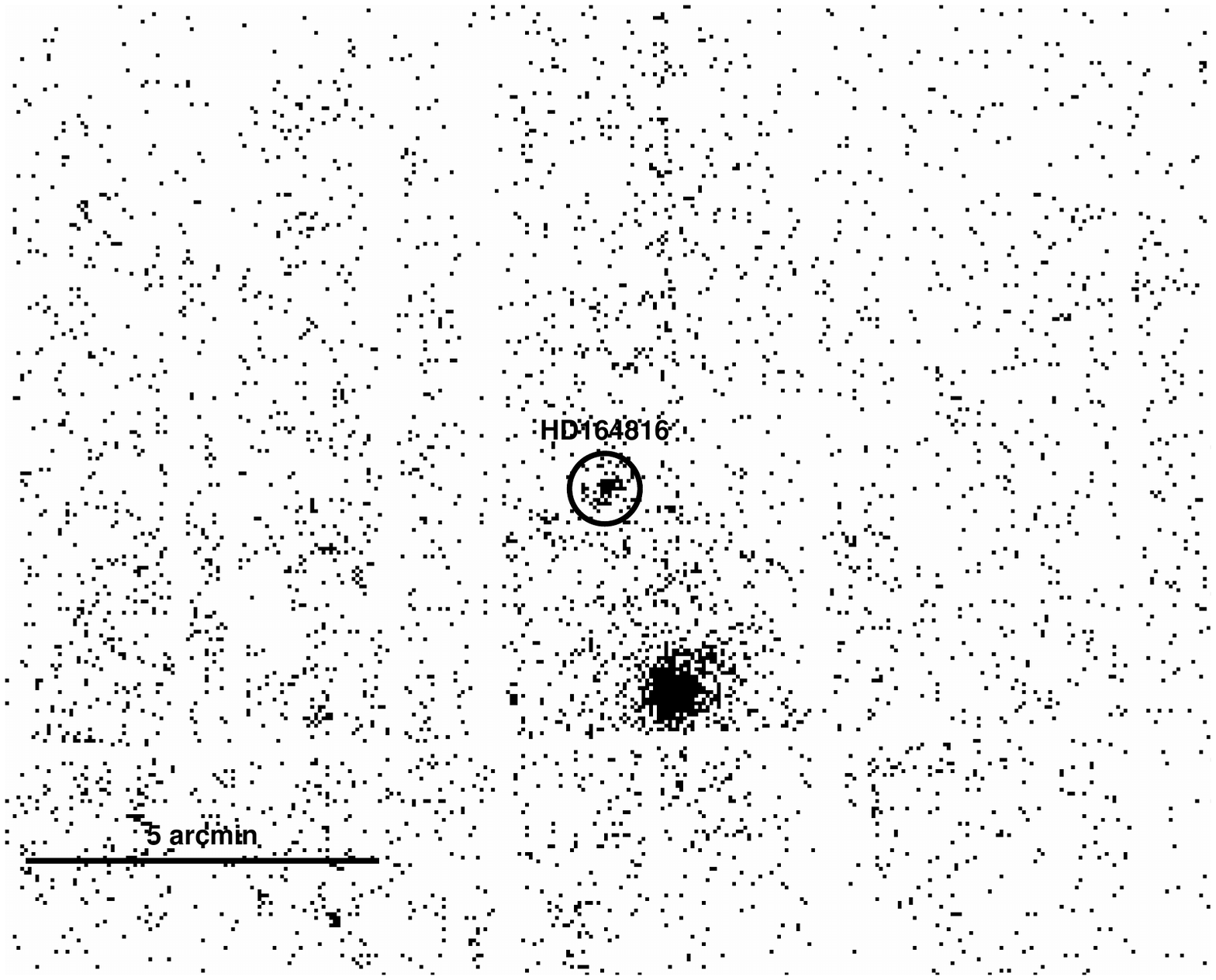}
  \caption{{\it Left Panel}: Image showing a 17$\times$14 arcmin$^{2}$ close-up view of the
           EPIC PN observation centered at the position of HD 164794 (9 Sgr) in the energy range 0.2 - 10.0 keV.
           Shown are the position of \hd~with the circle indicating the 30
           arcsec extraction region together with the background extraction
           region. As well seen is the rather large contamination by other sources
           in the field and the contribution of fringes from the
           optical emission of 9 Sgr. In addition surface plots of the extracted
           source (left inset) and background regions (right inset) are shown.
           {\it Right Panel}: Here we show \hd~in the energy range 0.2-0.5 keV with only single events displayed.}
  \label{hd_pn_ima}
\end{figure*}

We report here on the spectral and temporal analysis of the X-ray source \xmmssc~that is
coinciding with the optical position of \hd~and has been found by our search for
pulses in X-ray sources near galactic O-type stars.\\
We extracted events from the archived pipeline produced \emph{XMM PN} data (see table \ref{X_obs})
in a circular region with a radius of 30 arcsec around the centroid of \xmmssc~located
approximately 3.1 arcmin from the pointing of the observation by using the XMM Science Analysis System (SAS)
version 10. 
As background we extracted events in a nearby source free region of similar size. We used only
single-pixel event types for the spectral analysis as well as for the performed source detection
in the first energy band of \emph{XMM-Newton} (0.2 - 0.5 keV) as there are known calibration issues. Event types higher
than single-type are known to result in an elevated soft X-ray background which is most prominent in the reported first
energy band\footnote{Please see document XMM-SOC-CAL-TN-0068 on
http://xmm2.esac.esa.int/external/xmm\_sw\_cal/calib/documenta\-tion/index.shtml for reference}.\\
Extracting the background region is a somewhat complicated task as the field is very crowded and contaminated by
several fringes. Although the thick filter was in place in the \emph{XMM} Observation, those might result from 
incomplete blocking of UV/optical light coming from the 5th magnitude star 9 Sgr
(see Fig.\ref{hd_pn_ima}) or possibly originating from stray light of the nearby Low Mass X-ray
binary (LMXB) GX 9+1 (Langmeier, Sztajno, Tr\"umper \& Hasinger 1985) around 1 deg
outside the field of view (fov). Checking the background light curve of this observation
we could not identify any times of elevated background and thus can use the
whole observation time as a good time interval. Nevertheless we also investigated the
spatial distribution of the arriving photons in the background extraction region and
could not find any significant concentration, rather a homogeneous distribution (see inset
in Fig.\ref{hd_pn_ima} left panel).\\
The \emph{Chandra} observations do not show any fringes or an otherwise enhanced 
background as the threshold for optical contamination is $V \sim$ 3.1 for the used ACIS-I detector\footnote{Please see the 
most recent \emph{Chandra} proposers guide, vers 14.0 December 2011 for reference} (\hd~has  
$V$=7.09). Hence, the chosen background regions are not contaminated 
by any source or fringe.
\begin{table*}
\caption{X-ray observation log of the used \emph{XMM-Newton} and \emph{Chandra} observations}
\label{X_obs}
\begin{center}
\begin{tabular}{lccccc}
\hline
Observation ID & Pointing & Start Time & exposure time (ks) & Detector & offset from \hd~(arcmin) \\
\hline
\hline
 & & & & & \\
 & \emph{XMM-Newton} & & & & \\
0008820101 & HD164794 (9 Sgr) & 2001-03-08 11:21:27 & 23.572 & MOS1/2, PN & 3.07 \\
 & & & & & \\
\hline
 & & & & & \\
 & \emph{Chandra} & & & & \\
3754 & M8 & 2003-07-25 17:27:12 & 129.600 & ACIS-I & 4.27 \\
4397 & M8 & 2003-07-24 10:07:26 & 14.820 & ACIS-I & 4.27 \\
4444 & M8 & 2003-07-28 00:00:07 & 30.190 & ACIS-I & 4.27 \\
\hline
\end{tabular}
\end{center}
\end{table*}

\subsection{Spectral Analysis with \emph{XMM Newton}}
\begin{figure*}
  \centering
  \includegraphics[width=8.5cm, angle=180]{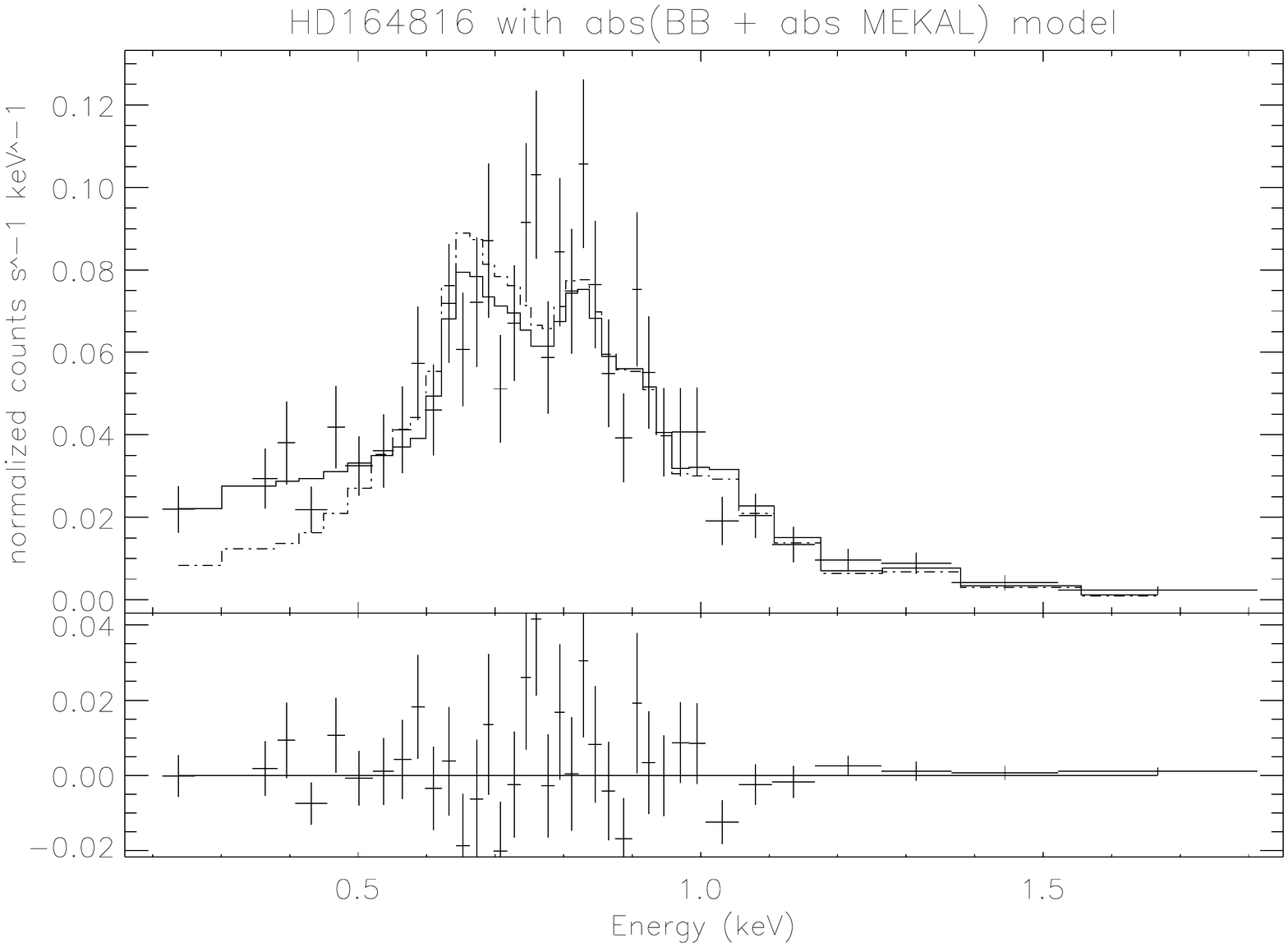}
  \includegraphics[width=8.5cm, angle=180]{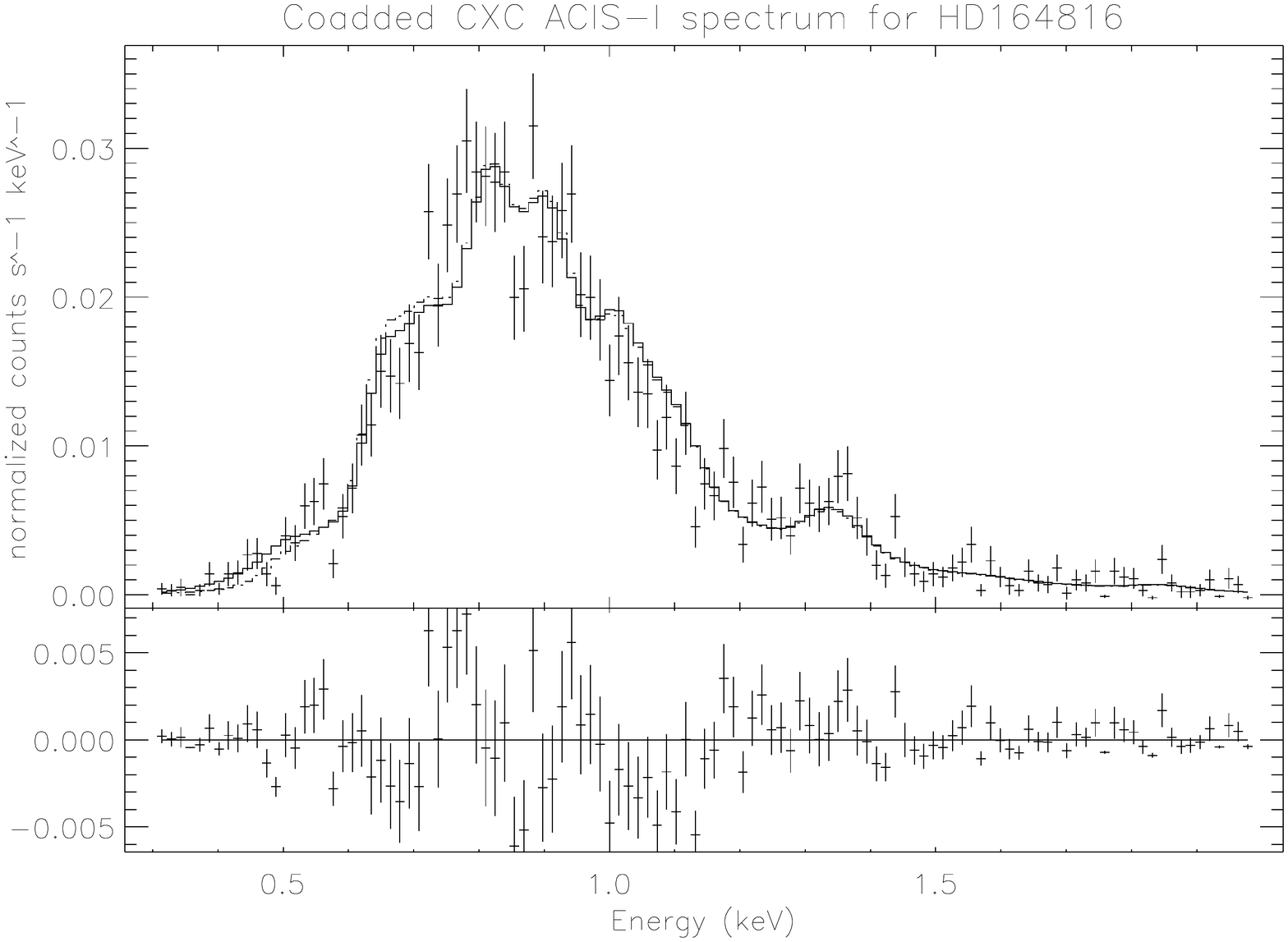}
  \caption{
           {{\it Left Panel}: Shown here is the best fit absorbed blackbody plus MEKAL model (continuous line) to
           the \emph{XMM PN} data together with the residuals for a blackbody
           temperature of kT$_{bb}\sim$48.79 eV. Data points from 0.27 keV to 0.32 keV are
           not taken into consideration in the fit due to the drastic dip in the effective area
           around 0.3 keV. The dash-dotted line represents the best fit absorbed MEKAL model to the data 
           and one can see the already mentioned excess from 0.2 to 0.5 keV.
           {\it Right Panel}: The coadded \emph{Chandra ACIS-I} observations are shown here
           together with the best fit absorbed blackbody + MEKAL plasma model and their
           corresponding residuals for a blackbody temperature of kT$_{bb}\sim$49.85 eV. Likewise the dash-dotted line 
           represents the best fit absorbed MEKAL model, where the excess from $ \sim$0.3 to $\sim$0.5 keV becomes 
           visible.}
          }
\label{hd_xmm_spec}
\end{figure*}

As a first step we have re-analyzed the X-ray spectrum of \xmmssc.
Its spectral properties have already been reported in Rauw et al. 2002 where
they performed an X-ray population study on an \emph{XMM Newton} observation and
among other things conclude that the X-ray emission of \hd~is typical for an
O9.5 III-IV star. However, they did not take the low energy part of the spectrum from
0.2 - 0.5 keV into consideration.\\
Trying to reproduce their results we first re-analyzed the \emph{XMM PN} dataset
of ObsID 0008820101 carried out with the thick filter in use and centered at
9 Sgr.\\
As a following step Response and Anxillary files have been computed by using
the XMMSAS tasks RMFGEN and ARFGEN. The data then have been grouped by a minimum
of 20 counts per energy bin in order to minimize the scattering of data points.
After background subtraction there are 961 net source counts available.
Including as well the first \emph{XMM~Newton} energy band (0.2 - 0.5 keV) in the spectral fit
by using a warm absorbed MEKAL plasma (Mewe et al. 1985, Kastra 1992) results in an
unsatisfying fit of $\chi^{2}_{\nu}$ equal 1.57 with 34 d.o.f which is not in agreement
with what Rauw et al. 2002 have found (i.e. $\chi^{2}_{\nu}$=1.08 in 0.5 - 5 keV).
By looking at the spectrum we clearly notice an excess in the range 0.2 - 0.5 keV that cannot be
modeled by the used absorbed MEKAL plasma model. In addition we recognize some
feature around 0.3 keV but we ignored data at the energy range 0.27 to 0.32 keV as there is
a drastic dip in the effective area of the \emph{EPIC PN} detector in that energy range.\\
We have carried out a source detection
in the energy range 0.2-0.5 keV in order to exclude an artificial nature of the detected low energy
excess. This energy part is believed to be mostly from the possible compact companion and has not been
regarded by Rauw et al. (2002). 
We detected \xmmssc~in this energy range with a
detection maximum likelihood of $L\sim$235 with $L=-\ln p$ where $p$ is equal to the probability that
the detected signal was generated by a random fluctuation. Since we only took single-pixel events into
consideration we can rule out that the excess is due to elevated background. As the given likelihood
value for the source detection can be considered significant we conclude that the excess is of
real nature (see right panel in Fig.\ref{hd_pn_ima}). A source detection in the energy range 
2.0 to 10 keV leads to a null detection, hence the source is soft.\
Improving the fit by blindly adding a blackbody component
could not be achieved thus we decided to use a more sophisticated approach. Therefore we tried
to subtract the MEKAL component from the spectrum.\\
In order to eliminate the thermal plasma component we used XSPEC (version 12.6) to simulate an
absorbed MEKAL plasma in the energy range 0.5-2.0 keV as there is negligible
information in the range 2.0-10.0 keV. As an input the best fit parameters for temperature
and absorption reported in Rauw et al. 2002 were used. Afterwards this simulated spectrum
is used to create a fake data set with the same exposure time, background, response
and anxillary files as used for the observation of \hd. The faked data are grouped according to
the \emph{EPIC PN} data (i.e. a minimum of 20 counts/bin). This
grouped MEKAL spectrum is then used as background for the observation in order to
 subtract the contribution of the thin plasma model.\\
Visually checking the produced spectrum confirms the MEKAL component to be consistent
with zero. The remaining excess is fitted with an absorbed blackbody model which leads
to a best fit temperature of around 53.7 eV that now serves as an initial input for
the future complete analysis.
As a next step the entire spectrum is fitted with an absorbed thin plasma model plus the blackbody component by
iteratively fixing and freeing the parameter pairs for the blackbody and MEKAL plasma model
(i.e. pairs of temperature and normalization) which lead us to a best fit blackbody temperature of
52.36$^{+7.78}_{-4.86}$ eV with $\chi_{\nu}^{2}$ = 1.06.\\
As the interstellar column density derived by Diplas \& Savage 1994, 0.15$^{+0.05}_{-0.04}\cdot$10$^{22}$ cm$^{-2}$,
is significantly smaller than the value derived by our fit (0.40$^{+0.03}_{-0.05}\cdot$10$^{22}$ cm$^{-2}$) we
decided to introduce a second systemic absorption component. Such a systemic absorption was already suggested
by Rauw et al. 2002 and could possibly be attributed to a stellar wind. By introducing such a component we can put tighter
constraints on the errors of the individual parameters and get an interstellar absorption of
0.08$^{+0.07}_{-0.03}\cdot$10$^{22}$ cm$^{-2}$ which is consistent with the one found by Diplas \& Savage 1994.
The resulting blackbody temperature is 48.79$^{+4.01}_{-8.65}$ eV (see Fig.\ref{hd_xmm_spec}).
Values for the systemic absorption and MEKAL temperature are 0.32$^{+0.09}_{-0.10}$$\times$10$^{22}$ cm$^{-2}$
and 0.24$\pm 0.02$ keV respectively. The absorbed flux in the energy range 0.2-2.0 keV of
1.08$\times$10$^{-13}$ erg cm$^{-2}$s$^{-1}$ is found to be higher than the
presented value for the energy range 0.5-5.0 keV by Rauw et al. 2002. For completeness we have computed as well the
unabsorbed flux in the 0.2-2.0 keV range (2.89$\times$10$^{-13}$ erg cm$^{-2}$s$^{-1}$). In addition we have used the
blackbody model normalization to give an estimate on the radius of the emitting area. Using a distance of 864$\pm$136 pc
results in a radius of the (circular) emitting area of 7.14$\pm$1.12 km which is in the range of a NS 
(see Hohle et al. 2012). 
Adopting the same distance range we 
used the MEKAL normalization to give an estimate on the emission measure (EM) of \hd, which turns out to be 
3.93$\pm$0.10$\times$10$^{54}$ cm$^{-3}$.  Those values are typical for O9V single stars 
(see e.g. Bhatt et al. 2010 and Naz\'e et al. 2011).
 An overview
of the most important parameters can be found in table \ref{spec_par}.

\subsection{Spectral Analysis with \emph{Chandra}}
\hd~has been in the field
of view of six \emph{Chandra} observations of which we unfortunately had to exclude
two as the high energy transmission grating (HETG) was in place and \hd~thus not
covered by the active part of the detector. The remaining four observations (ObsIDs
977, 3754, 4397, 4444; see table \ref{X_obs}) taken by the Advanced CCD Imaging Spectrometer 
in Imaging mode (ACIS-I) have 
been reprocessed individually, as follows. First we extracted events in a circular
region around \hd~with a radius of 10 arcsec for each observation and as there are
neither any fringes nor any other signs for an elevated background present we arbitrarily
chose one background field per observation with a radius of 20 arcsec in a source
free region in close proximity to \hd. We chose a somewhat bigger extraction region
for the background as the background level of \emph{Chandra} observations is in
general very low.\\
Response and Anxilary files have been created afterwards for source and background
regions each by using the CIAO (version 4.2) tools MKRMF and MKARF. As we noticed that the source
counts are low in the single observations we inspected the effective area
of each observation and found the differences for ObsIDs 3754, 4397 and 4444 to be
indistinguishable and hence combined these three event files. However for ObsID 977 we
find the effective area to be significantly smaller and thus did not include this
observation. Coadding the remaining three observations lead to a total effective
exposure time of $\sim$172 ks with 1203 counts available for \hd~and thus
having slightly higher statistics compared to the \emph{XMM} observation.\\
For the following spectral analysis we applied the same methods and constraints as to the
\emph{XMM PN} observation but we used the C-statistic instead of $\chi^{2}$ in that case.
As a best fit blackbody temperature we get 49.85$^{+1.29}_{-8.75}$ eV with $C$ = 168.77 for 109
d.o.f. in the energy range of 0.3-2.0 keV (see Fig.\ref{hd_xmm_spec} right panel), as the
ACIS-I detector is only calibrated from 0.3 to 11.0 keV. Comparing the MEKAL plasma temperature
and both absorption components obtained from the \emph{Chandra} data with the values from
\emph{XMM PN} data we note that they are in agreement. This is also true for the values of the
MEKAL plasma temperature in the Rauw et al. 2002 analysis.
The computed absorbed and unabsorbed flux are 1.19$\times$10$^{-13}$ erg cm$^{-2}$s$^{-1}$
and 2.19$\times$10$^{-13}$ erg cm$^{-2}$s$^{-1}$ respectively. Computing the radius of the emitting area 
864$\pm$136 pc we end up with R$_{emit}$ = 7.07$\pm$1.11 km which is consistent with the value derived from the XMM fit. 
By using the MEKAL normalization factor we find the EM in the range 
3.76$\pm$0.09$\times$10$^{54}$ cm$^{-3}$, which is similar to the XMM data.\\
We summarize our X-ray results in Table \ref{spec_par}.
\begin{table*}
\caption{Spectral parameters inferred from fitting the \emph{Chandra} and \emph{XMM-Newton} spectra of
\hd.}
\begin{center}
\begin{tabular}{lccccccc}
\hline
\hline
Model $^{a}$ & $\chi^{2}$ / $C-$stat & D.O.F. & interstellar $n_{H}$ / systemic $n_{H}$ & $kT_{MEKAL}$ & $kT_{BB}$ & $R_{emit}$  & $f_{X}$ (0.2-2.0~keV) \\
      &                  &        & ($10^{22}$ cm$^{-2}$) & (eV)  & (eV) & (km) & ($10^{-13}$ erg cm$^{-2}$ s$^{-1}$)\\
      \hline
& & & \emph{XMM-Newton} & (0.2 - 2.0 keV) & \\
\hline
\\
abs*MEKAL   & $\chi^{2}=53.30$  & 34 & $0.35^{+0.08}_{-0.07}$ & $237.81^{+19.61}_{-11.48}$  & --- & --- & $7.90$  \\
\\
abs(BB+ abs*MEKAL)   & $\chi^{2}=32.84$  & 31  & $0.08^{+0.07}_{-0.03}$ / $0.32^{+0.09}_{-0.10}$ &  $237.81^{+23.44}_{-19.68}$ &$48.79^{+4.01}_{-8.65}$ & 7.14$\pm$1.12 & $2.89$  \\
\\
\hline
Model $^{a}$ & $\chi^{2}$ / $C-$stat & D.O.F. & interstellar $n_{H}$ / systemic $n_{H}$ & $kT_{MEKAL}$ & $kT_{BB}$ & $R_{emit}$  &$f_{X}$ (0.3-2.0~keV) \\
      &                  &        & ($10^{22}$ cm$^{-2}$) & (eV)  & (eV) & (km) & ($10^{-13}$ erg cm$^{-2}$ s$^{-1}$)\\
\hline
& & & \emph{Chandra} & (0.3 - 2.0 keV) & \\
\hline
\\
abs*MEKAL   & 184.68  & 112  & $0.39^{+0.01}_{-0.02}$ & $227.70^{+10.40}_{-5.06}$  & --- & --- & $1.01$  \\
\\
abs(BB+ abs*MEKAL)   & 168.77  & 109  & $0.09^{+0.11}_{-0.09}$ / $0.28^{+0.09}_{-0.16}$  &  $238.90^{+12.52}_{-10.77}$ & $49.85^{+12.92}_{-9.56}$ & 7.07$\pm1.11$ & $2.19$  \\
\\
\hline
\hline
 \end{tabular}
 \end{center}
 $^{a}$   {\footnotesize BB = blackbody; MEKAL = hot diffuse gas model based on Mewe et al. 1985, 1986, Kaastra 1992 and Liedahl et al. 1995 }\\
\label{spec_par}
\end{table*}

\subsection{X-ray pulsation search with \emph{XMM Newton}}
\begin{figure}
  \centering
  \includegraphics[width=8.5cm, angle=180]{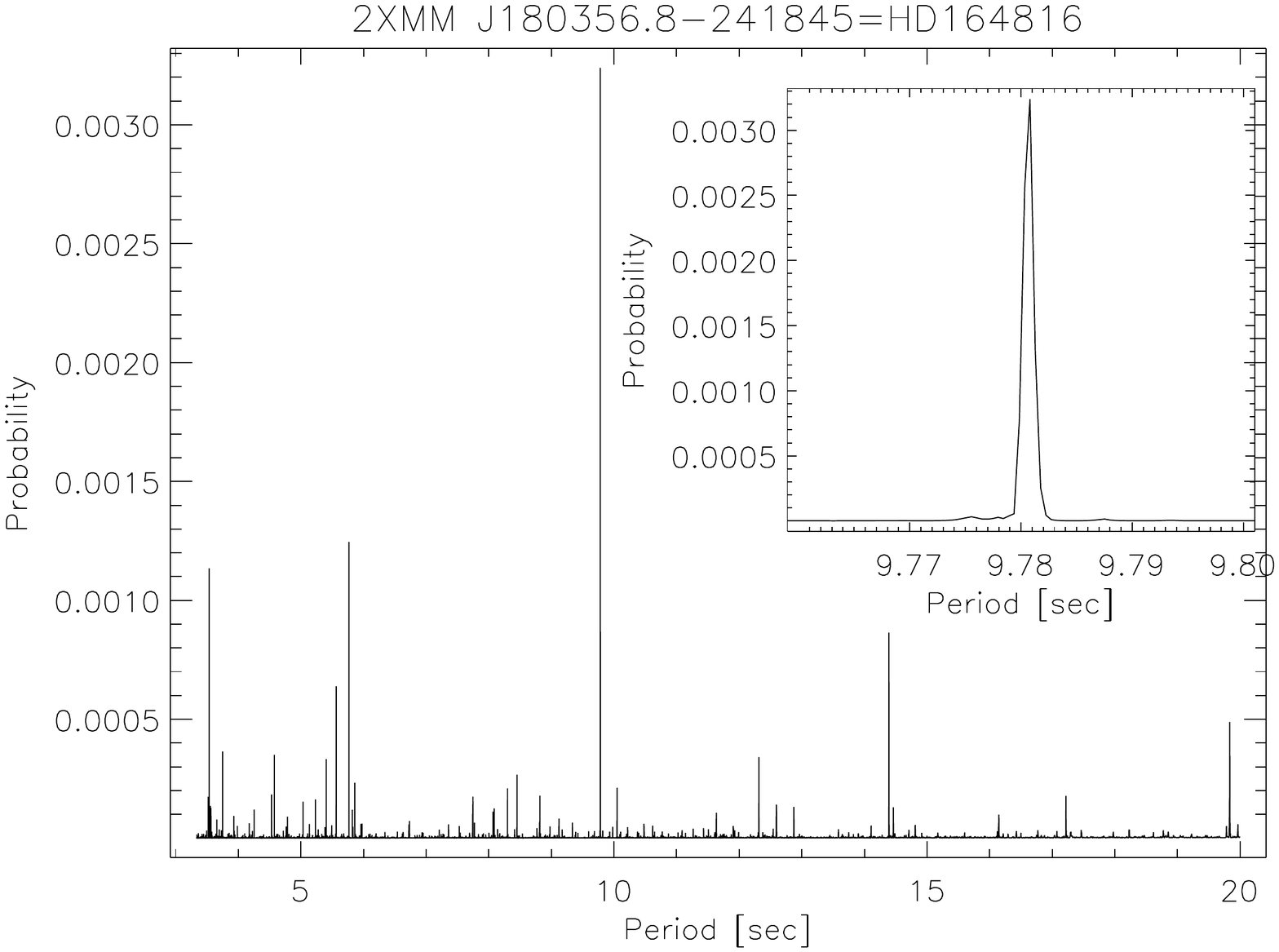}
  \caption[]{Here we show the periodogram of the detected $\sim$9.78 s pulsation for the 
             X-ray source identified with \hd~found by the bayesian approach with \emph{XMM Newton}.}
  \label{bay_per}
\end{figure}

We performed a periodicity search at 
the position of
\xmmssc. For performing the pulsation search only the \emph{XMM Newton} observation has been reprocessed
as the \emph{EPIC PN} detector was used which is the only available
instrument for this source  with a sufficiently short readout time (73 ms in that configuration).
In particular there are 1284 arrival times in this set that have been searched for a periodic signal
by means of the bayesian approach developed by Gregory \& Loredo 1992 for the detection of a signal
with unknown shape. Applying this method leads to a 
detection of a signal at 
9.7804$^{+0.0007}_{-0.0003}$ s (see Fig.\ref{bay_per}) with a probability that the data favor a
periodic model over a constant model to be 0.56. This resulting period could afterwards be reproduced
by applying the Z$_{m}^{2}$ test with m being the number of harmonics (Buccheri et al. 1983) in the period range 1 to 10s 
with a stepsize equal to the independent fourier spacing (IFS) i.e. 1/T$_{span}$. This
number of harmonics has been optimized by using the H-Test (De Jager, Swanepoel \& Raubenheimer 1989)
and  was found to be maximized for m=1. The resulting Z$_{1}^{2}$ value is equal to 22.67 at a
period of 9.7813$^{+0.0005}_{-0.0007}$ s with the number of expected peaks exceeding that Z value 
to be 5.25$\times$10$^{-6}$. The given errors are 1 $\sigma$ errors. Finally a pulsation search with a method 
based on the cash statistic (Cash~1979) as described in Zane et al. (2002) has been applied to the arrival times 
and could determine the period to be 9.78$\pm 0.06$ s (again in the range 1 to 10s and using the IFS). 
The errors are again at the 1 $\sigma$ level.\\
Hence, the period found by the bayesian method, being the proper choice
for this kind of data as it uses an unbinned approach, is confirmed by the cash and Z$_{1}^{2}$-test. In addition 
we have created a simulated data set of equal length in exposure time with 1284 randomly 
distributed photons. Applying the Z$_{1}^{2}$-Test and the cash-test to that set of data lead to no detection at 
$\sim$9.78 s. Furthermore the most likely periods found by those two tests in the faked data set are not in accordance 
(P$_{Z_{1}^{2}}\sim$5.15 s and P$_{cash}\sim$1.12 s).\\
The CXC ACIS-I data sets although improving the number of source photons have not been used
for timing analysis as the ACIS frame time is $\sim$3.24 s and thus too long to significantly detect such a
period.\\
\begin{figure}
  \centering
   \includegraphics[width=8.5cm, angle=180]{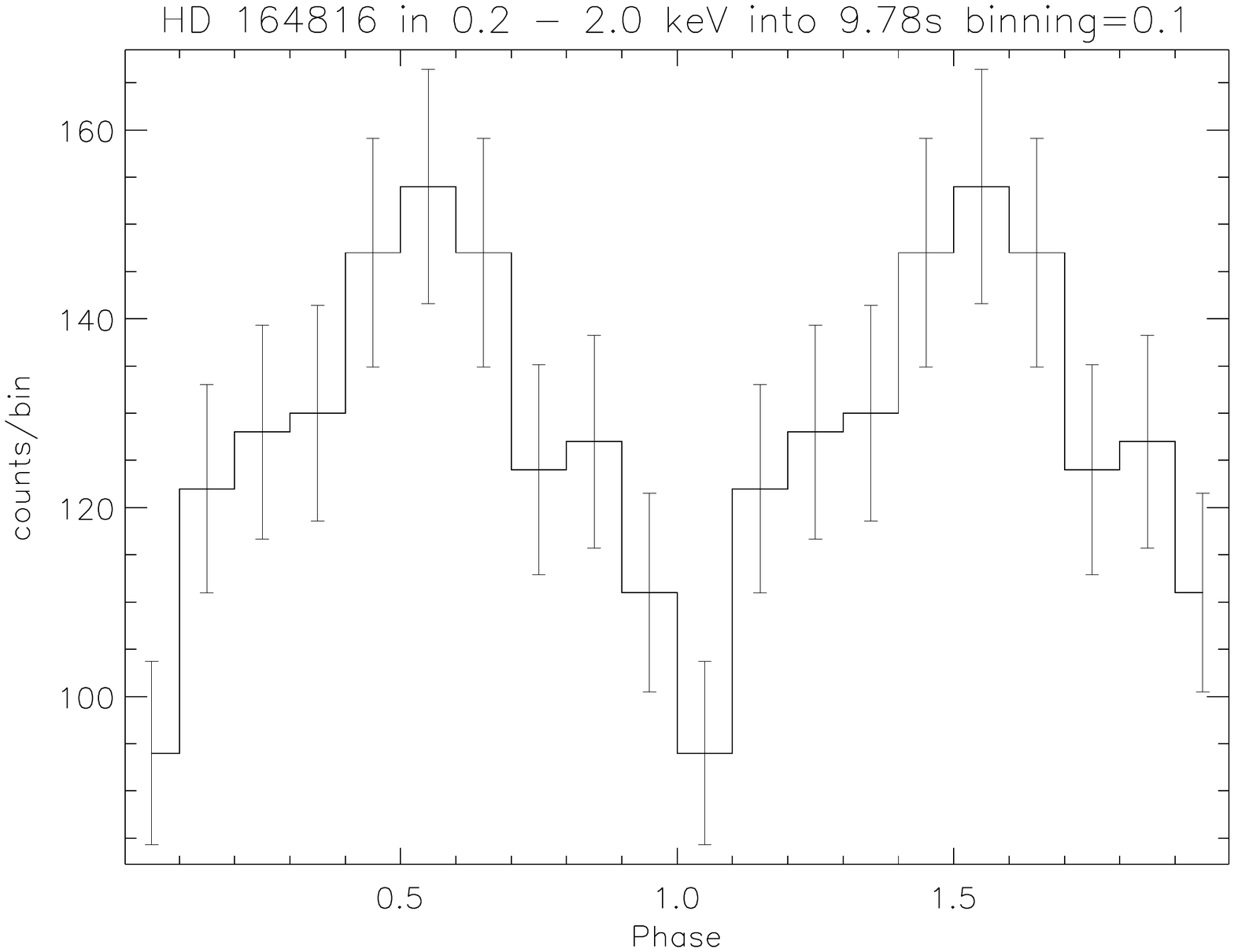}
   \caption{In this figure we show the phase-folded light curve for the X-ray source in the energy range 0.2-2.0 keV 
            folded into the resulting best period of 9.78 s resulting from \emph{XMM Newton} data. For clarification 
            we show two full cycles with a binning of 0.1.}
\label{compact_ltcrv}
\end{figure}
The phase-folded X-ray light curve for \xmmssc~in the energy range 0.2-2.0 keV has a pulsed
fraction of 60.4$\pm$15.4 \% (c.f. Fig.\ref{compact_ltcrv}). 
As a final step we extracted arrival times for the brightest X-ray source in the field of view
(i.e. the O4 star 9 Sgr) in the energy range 0.2-2.0 keV which results in 18769 counts. We folded the
created light curve with the obtained period which results in a constant non-periodic signal.
This is likewise true for the extracted background region, again in the energy range 0.2-2.0 keV.
Considering this non-periodic signal of the background and of 9 Sgr, 
we conclude that the observed period in \hd~
is not due to some background modulations caused for example by the readout time or multiples
of it.\\
Both the soft excess and the $\sim$10 s pulsation would be typical for a NS in the system 
assuming that it will not be interacting with its host star and can hence be treated as effectively isolated 
(see Haberl 2007 for typical periods of thermally emitting isolated NS), namely $\sim$3 to $\sim$11 s.

\section[]{Orbital parameters of \hd}
\subsection{Optical data analysis}
Optical \'echelle spectroscopy of \hd~ has been obtained at two observatories
in Chile: the Cerro Armazones Observatory (OCA, about 20 km east of the Paranal
Observatory) and at the Cerro Tololo Interamerican Observatory (CTIO).\\
Optical spectroscopy at the OCA was secured using the Bochum Echelle Spectrographic Observer
(BESO) fiber-fed from the focus of the 1.5m Hexapod Telescope (see Fuhrmann et al. 2011).
Fifteen spectra were obtained between 2009-04-21 and 2010-03-28 covering the
wavelength range of 3530 - 8860~\AA~ with a spectral resolution of $R$ = 50000 (Fuhrmann et al. 2011) 
(see Table \ref{journal.tab}). The data were reduced using dedicated scripts
written under MIDAS. The reduction includes overscan, bias and
flatfield correction; individual \'echelle orders were extracted, wavelength
calibrated and normalized to the continuum. Finally, cosmic spikes have been removed.
The cleaning was complicated by the fact that most He and H lines contain narrow
(and variable) emission features. Moreover, for each night there is just
one exposure. Wavelength calibration has been improved using telluric bands
close to 6900 and 7600~\AA~and the spectrum of $\eta$ CMa as a template. The
improved radial-velocity system is stable to about 100 m/s. The zero point of the radial-velocity system 
was checked by measuring the IAU standard star HIP 910, which is a slow rotator, in 6 nights. The average 
radial-velocity is 14.16$\pm$0.13 km/s which is in good agreement with 14.4$\pm$0.9 km/s found by Evans (1967).\\
Optical spectra at the CTIO were obtained using the high dispersion optical Blanco \'echelle
spectrograph fiber-fed from the 1.5m SMARTS telescope. Fourteen spectra taken between
2010-07-25 and 2010-09-14 cover the range from 4820 - 7120~\AA~(see Table \ref{journal.tab}).
The $R\approx$20000 spectra
were extracted using software written in IDL \footnote{http://www.astro.sunysb.edu/fwalter/SMARTS/ech\_proc.txt}.
The orders are traced using the cross-dispersed flat images. The spectrum is extracted
using a boxcar extraction, as is the flat spectrum. The spectrum is then divided
by the extracted flat spectra. In general we obtained 3 spectra at each epoch.
These are scaled, median-filtered to reject cosmic rays, and summed.
Wavelength calibration is based on a Th-Ar calibration lamp exposure taken
just prior to each stellar observation. The zero-point of the wavelength scale
is uncalibrated at the level of 1 pixel (3 km/s). No attempt has been made to
convert to a flux intensity scale.

\begin{figure}
\centering
 \includegraphics[width=9.5cm, angle=90]{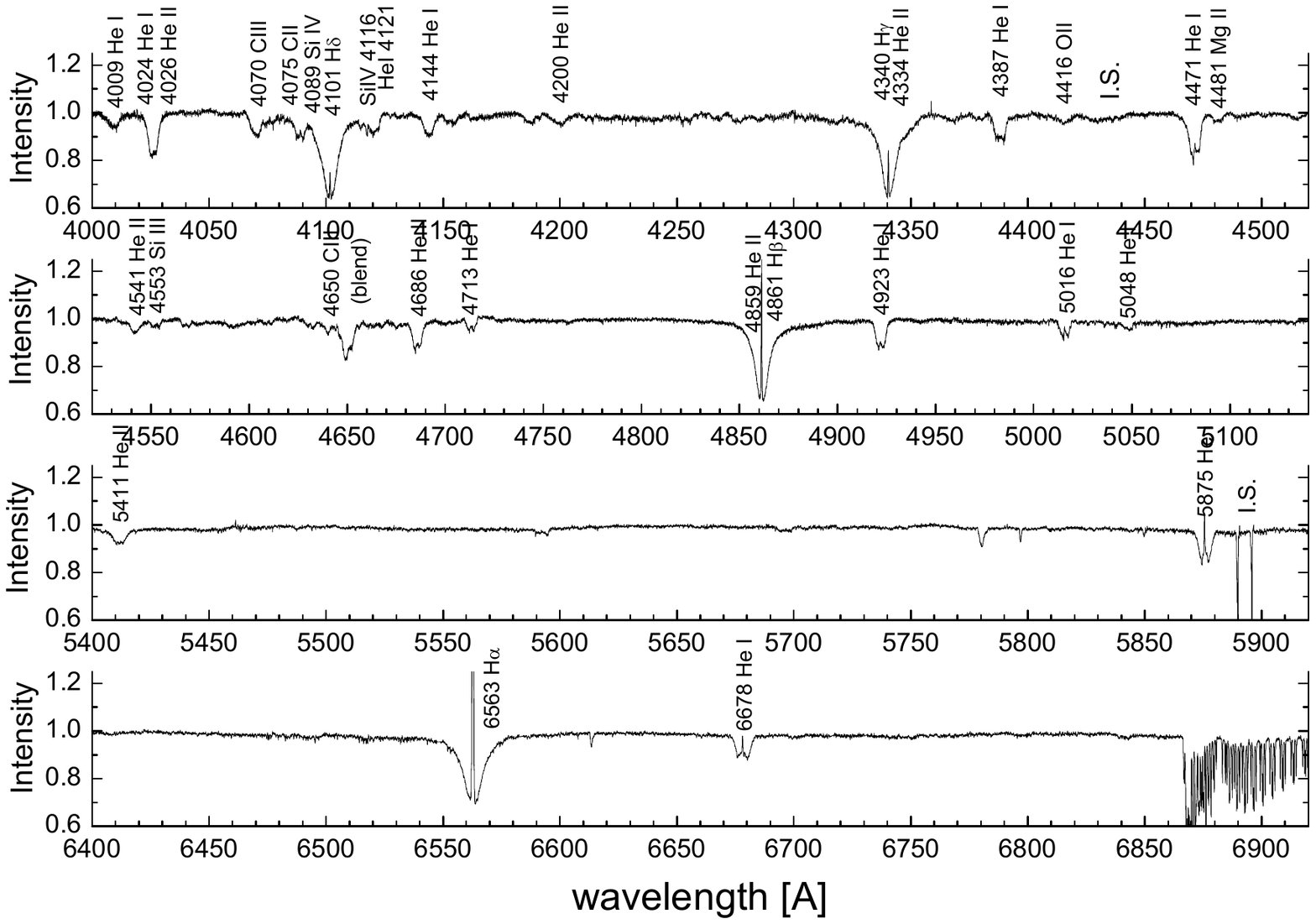}
 \caption{Average spectrum of all OCA observations on HD 164816 with line
           identifications (I.S. stands for interstellar band/line)}
\label{whole_spec}
\end{figure}

\begin{figure}
\centering
 \includegraphics[width=9.5cm]{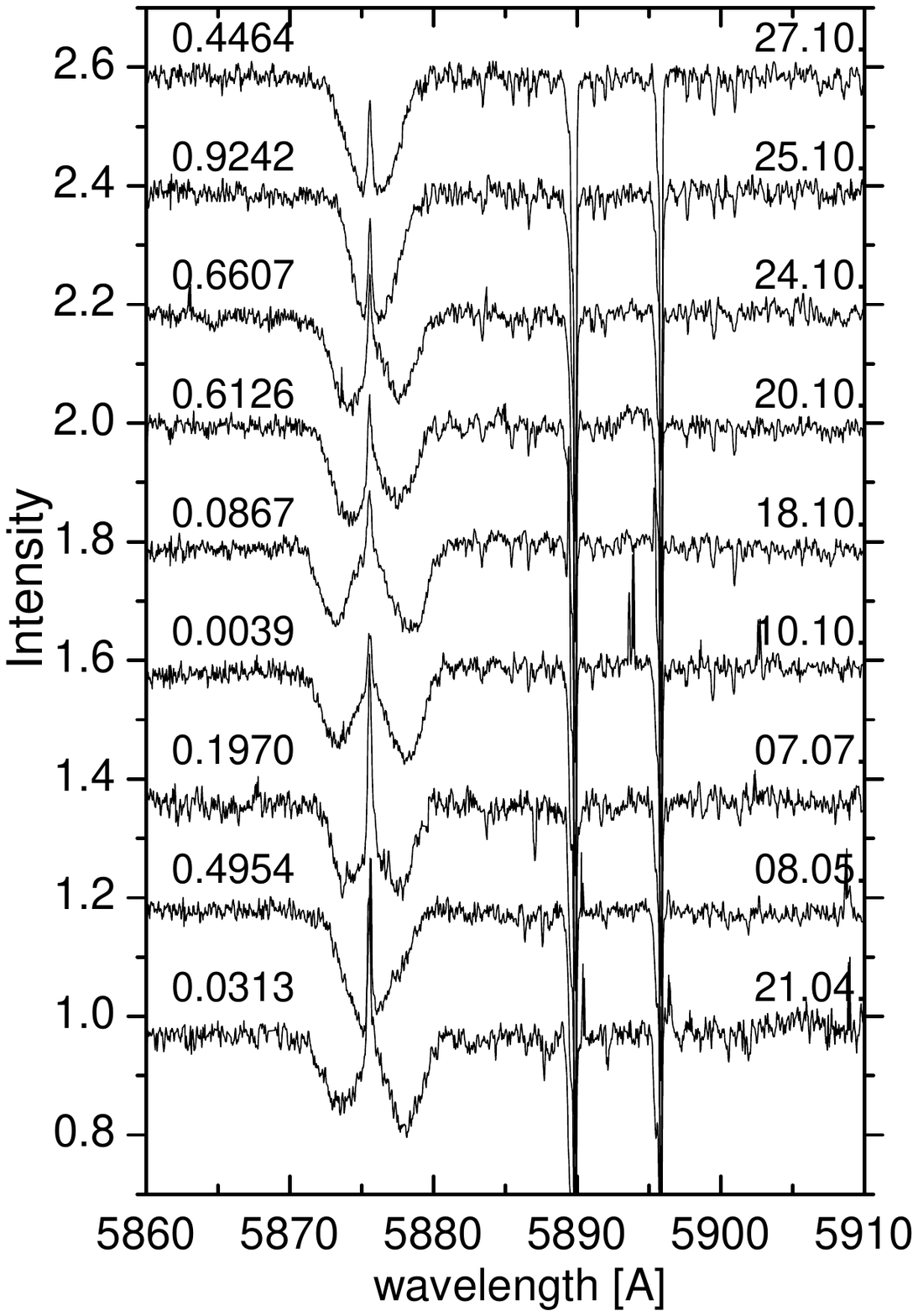}
 \caption{The He I 5876 \AA~ line profile changes with the orbital revolution of
          the binary while the interstellar neutral sodium doublet ($\lambda\lambda$
          5893, 5896\AA) is not following the orbital revolution. The same holds for 
          the narrow nebula emission at 5876 \AA~ HeI, which is variable in strength but 
          constant in wavelength, i.e. radial velocity. The dates of observations
          (in 2009) and orbital phases (counted from the periastron passage) are also shown.}
\label{sb2_changes}
\end{figure}

\begin{table}
\caption{Journal of spectroscopic observations of \hd~obtained at the CTIO and
         OCA observatories
         \label{journal.tab}}
\begin{scriptsize}
\begin{center}
\begin{tabular}{lcr|lcr}
\hline
 Spectrum      & HJD        & Exp.  & Spectrum      & HJD        & Exp.  \\
               &2\,400\,000+& [sec] &               &2\,400\,000+& [sec] \\
\hline
 OCA\_20090414  & 54936.8391 &1800 &  OCA\_20100328   & 55284.8365 &1500 \\
 OCA\_20090421  & 54943.8225 &1800 &  CTIO\_20100725  & 55403.7934 & 200 \\
 OCA\_20090508  & 54960.8685 &1800 &  CTIO\_20100728  & 55406.7316 & 300 \\
 OCA\_20090707  & 55020.8242 &1800 &  CTIO\_20100731  & 55409.6988 & 300 \\
 OCA\_20091010  & 55115.5480 &1800 &  CTIO\_20100804  & 55413.6327 & 300 \\
 OCA\_20091018  & 55123.5011 &1800 &  CTIO\_20100805a & 55414.6293 & 300 \\
 OCA\_20091020  & 55125.5094 &1800 &  CTIO\_20100805b & 55414.7672 & 300 \\
 OCA\_20091024  & 55129.5112 &1800 &  CTIO\_20100812a & 55421.6210 & 600 \\
 OCA\_20091025  & 55130.5172 &1800 &  CTIO\_20100812b & 55421.7031 & 600 \\
 OCA\_20091027  & 55132.5113 &1800 &  CTIO\_20100820  & 55429.6515 & 400 \\
 OCA\_20100321  & 55277.9383 &1500 &  CTIO\_20100904  & 55444.5716 & 600 \\
 OCA\_20100323  & 55279.9109 &1500 &  CTIO\_20100907  & 55447.5156 & 600 \\
 OCA\_20100324  & 55280.8909 &1500 &  CTIO\_20100908  & 55448.5887 & 600 \\
 OCA\_20100325  & 55281.8479 &1500 &  CTIO\_20100909  & 55449.5630 & 600 \\
 OCA\_20100327  & 55283.8057 &1500 &  CTIO\_20100914  & 55454.5746 & 600 \\
\hline
\hline
\end{tabular}
\end{center}
\end{scriptsize}
\end{table}
Although the zero point of the radial-velocity system is uncertain, there is
very good agreement between the systemic (mass-center) velocity obtained
separately from the individual datasets (OCA and CTIO data).\\
Optical spectra of \hd~do not contain many features (see Fig.~\ref{whole_spec}):
in addition to the interstellar lines there are only lines of the Balmer series
and those of neutral and ionized helium and possibly faint lines of CII, CIII,
Si IV, and O II. In the OCA data forbidden lines, e.g., [OIII] at 5006.84 \AA,
are also visible, most likely residual features from the M8 nebular emission 
remaining after the background subtraction. The Balmer series
lines (and also some He I lines) show narrow and variable nebula 
emission lines always at the same wavelength (e.g. 5876 \AA (see Fig. \ref{sb2_changes})).\\
Our new spectra confirm the binarity (Penny 1996, Howarth et al. 1997, Mason et al. 2009) 
of the system (see Fig.~\ref{sb2_changes}).
We extracted the Doppler information and determined the spectroscopic
orbit\footnote{In the case of the CTIO spectra only 4923 \AA, 5016 \AA, 5875 \AA,
6678 \AA~lines were covered}; we used nine spectral regions centered at sufficiently
strong He I/II lines ($\lambda\lambda$ 4024/4026, 4387, 4471, 4686, 4713, 4923, 5016,
5875, 6678 \AA). The strong forbidden line [OIII] 5006.84 \AA~has been removed prior to
the fitting (visible mainly in the OCA data). \\
The global modeling of the data included orbital parameters ($P$, $e$, $\omega$, $\rm{T_0}$, $V_0$,
$K_{1,2}$), relative intensities ($I_{1,2}$) and projected
rotational velocities of the components ($v_{1,2} \sin i$); see Table~\ref{opt_par} for
the individual parameters. Prior to the fit all emission seen in He lines has been cleaned 
"manually" in order to secure a proper determination of the radial-velocity. The fitted spectrum 
consisted of the sum of two appropriately broadened and Doppler shifted synthetic spectra for each
orbital phase. Theoretical rotational profiles (see Gray, 1976) were computed
assuming a linear limb darkening law and solid-body rotation.
For $T_{eff}$ = 32500 K, $\log g$ = 4.00 [log(cm~s$^{-2}$)] and the wavelength range of the modeled He lines
the linear limb darkening coefficient is $0.20 < u < 0.28$ (see van Hamme, 1993). Because
the shape of the rotational profile depends on the limb darkening only slightly, an average
value of $u$ = 0.24 was used. The template high-resolution synthetic
spectrum taken from the Pollux database \footnote{http://pollux.graal.univ-montp2.fr/}
corresponding to $T_{eff}$ = 32500 K, $\log g$ = 4.00 and solar metallicity
provided a very good fit to the data. Synthetic spectra corresponding to higher
temperatures showed helium lines being partially in emission. On the other hand,
a synthetic spectrum corresponding to 30000 K resulted in smaller equivalent
widths for the helium lines as observed. It is also possible that the helium
abundance is higher than assumed in the synthetic spectra while the temperature
is slightly lower. Previous determinations of the spectral type of HD164816
are mostly B0V (see e.g. Morris 1961). Our spectroscopy indicates O9V or O9.5V as the best estimates.\\
In the first step we tried to find the approximate orbital period for the system
 using the full width at half maximum changes 
of the strong He I line at 5875 \AA. The present set of
29 spectra spanning 17 months indicated only one possible orbital period,
$P \sim 3.82$ days. First fitting experiments, however, showed it was impossible
to satisfactorily fit all the data without the assumption of a fast apsidal motion.
Hence we added $d\omega/dt$ to the parameter set with $\omega_0$ valid for
the periastron passage $T_0$ (at an arbitrarily selected epoch of the binary).\\
Because the long-term All-Sky Automated Survey (ASAS\footnote{Please refer to
http://www.astrouw.edu.pl/asas/}) light curve (Pojmanski, 1997 \&  1998)
does not show eclipses, a constant relative intensity of the components throughout orbital
phases has been assumed. The best global fit to 29 spectra and 9/4 (OCA/CTIO) spectral
regions resulted in (among other parameters) $P$ = 3.81932(39) days and $e$ = 0.232(16). 
The reduced $\chi^2$ = 1.394 indicates a small inconsistence of the model and the fitted spectra. 
The larger $\chi^2$ very probably arises from the simple assumption of spherical shapes for the components and
also variations in the continuum rectification level.
\begin{table}
\caption{Spectroscopic elements inferred from fitting the 29 OCA/CTIO spectra.
         The 1$\sigma$ error of the last digit is given in parentheses. Reduced
         $\chi^2$ is given
         \label{opt_par}
         }
\centering
\begin{tabular}{lcc}
\hline
\hline
Parameter        &               &                   \\
\hline
$P$              & [days]        & 3.81932(39) \\
$e$              &               & 0.232(16)   \\
$\omega_0$       &  [rad]        & -0.97(7)    \\
$V_{\gamma}$     &  [km~s$^{-1}$] & -77.1(13)         \\
$K_1$            & [km~s$^{-1}$] & 109.4(26)   \\
$K_2$            & [km~s$^{-1}$] & 120.3(30)   \\
$T_0$            & [HJD]         & 2\,455\,000.88(3) \\
$d\omega/dt$     & [rad/yr]      & 0.67(7)     \\
$U$              & [yr]          & 9.4(11)     \\
$I_1$            &               & 0.536(14)   \\
$v_1 \sin i$     & [km~s$^{-1}$] & 85.4(28)    \\
$I_2$            &               & 0.463(14)   \\
$v_2 \sin i$     & [km~s$^{-1}$] & 79.9(32)    \\
$m_1 \sin^{3} i$     & [M$_\odot$]   & 2.31(17)    \\
$m_2 \sin^{3} i$     & [M$_\odot$]   & 2.10(16)    \\
$a \sin i$       & [R$_\odot$]   & 16.8(2)     \\
$\chi^2_{\rm r}$ &               & 1.394       \\
\hline
\hline
\multicolumn{3}{p{0.4\textwidth}}{The following elements were optimized: $P$ - period, $e$ - eccentricity,
           $\omega_0$ - longitude of periastron passage at $T_0$, $V_0$ - systemic
           velocity, $K_{1,2}$ - semi-amplitudes of radial velocity changes,
           $T_0$ - time of the periastron passage, $d\omega/dt$ - apsidal motion
           rate ($U$ is apsidal-motion period), $I_{1,2}$ - relative intensities of
           spectra, $v_{1,2} \sin i$ - projected rotational velocities of components. Projected
           masses of the components, $m_{1,2} \sin^3 i$, and projected
           major axis, $a \sin i$ are also given}
\end{tabular}
\end{table}

The resulting projected masses of the components $m_1 \sin^3 i$ = 2.31(17) M$_\odot$
and $m_2 \sin^{3} i$ = 2.10(16) M$_\odot$ indicate a low inclination angle for the system,
which is supported by the lack of eclipses. The  ZAMS mass of an O9.5V star (X = 0.70,
Y = 0.28, Z = 0.02; see Claret 2004) is about 16 $M_\odot$; the ZAMS radius is about 5 R$_\odot$
(at the cluster age of 2.3 Myr (Prisinzano et al. 2005) the star would have a corresponding radius of about 5.6 R$_{\odot}$). Then the inclination
angle would be 30 - 35\degr. Assuming $i$ = 30\degr and $a \sin i$ = 16.8(2) R$_{\odot}$
we get a semi-major axis of 33.6(4) $R_\odot$. In the case that the components are
rotating synchronously (the synchronization time scale is usually two orders of magnitude shorter
than circularization time scale; see e.g., Pan et al. 1998) their radii (using
$v_{1,2} \sin i$) are then $R_1$ = 13.2 R$_\odot$ and $R_2$ = 12.1 R$_\odot$,
indicating that both components already left the main sequence or that they are still rotating
asynchronously (with rotation factors $F_{1,2} \sim 2$). Without knowledge of
the inclination angle and in the view of possible asynchronous rotation of the
components determination of the true radii and masses is also complicated.

\subsection{Distance}
The individual distance towards HD 164816 was determined by Megier et
al. (2009) from interstellar Ca II absorption to be $864 \pm 136$ pc,
while the NGC 6530 cluster is at $1543 \pm 345$ pc (see Sect. 1). Since these
two values are deviant by more than $1~\sigma$, we will discuss the distance
of HD 164816 and whether it is or was a member of the NGC 6530 cluster.

Since the radii of the two O9.5 stars in HD 164816 could not be determined
directly by us, we cannot compute the distance by the Stefan-Boltzmann law
from temperature, radius, and luminosity. However, since the two stars are
located close to the dwarf sequence (V), we can compute the distance modulus
from the apparent magnitude of HD 164816 (corrected for both extinction and
binarity) and the typical absolute magnitude of an O9.5V star.

The published magnitudes of HD 164816 are (all in mag from Simbad and
references therein) U = 6.22, B = 7.09, V = 7.09, I$_{\rm C}$ = 6.99, J = 7.006,
H = 8.053, and K = 7.072 (BV from Hipparcos, U from Reed et al. 2003,
I Cousins from Rauw et al. 2002, JHK from 2MASS, Cutri et al. 2003),
all having small errors of roughly $\pm 0.01$ mag for UBVI and
$\pm 0.025$ mag for JHK.

Using the intrinsic UBVI$_{\rm C}$JHK colors of an O9.5V star
according to Bessell et al. (1998) for temperatures of $31250 \pm 150$ K
and $\log g = 5.0$ (for main sequence dwarfs), we can obtain the
extinction by comparing the apparent and intrinsic colors;
for the interstellar extinction law, we interpolate in each band
according to Cardelli et al. 1989, Savage \& Mathis 1979,
and Rieke \& Lebofsky 1985. Then, we obtain
A$_{\rm V} = 1.10 \pm 0.05$ mag as interstellar extinction towards HD 164816.

For the total bolometric luminosity of an O9.5V star, we use the latest
determination from Hohle et al. (2010) using Hipparcos distances
and extinction corrections from Hipparcos BV and 2MASS JHK colors,
all corrected for multiplicity, interpolating between O9V and B0V.
For the bolometric correction, we use again Bessell et al. (1998) for
temperatures of $31250 \pm 150$ K and $\log g = 5.0$ (for main sequence dwarfs),
namely B.C.$_{\rm V}=3.025 \pm 0.085$ mag.

Then, we obtain as (main-sequence spectro-photometric) distance towards \hd~
the value $1030 \pm 230$ pc. This value is consistent with the value by
Megier et al. (2009) from interstellar Ca II absorption being $864 \pm 136$ pc,
Hence, our assumptions appear justified. For O9IV, O9V, B0IV and B0V stars, the values 
lie between $890 \pm 40$ pc and $1240 \pm 140$ pc, i.e. are consistent within the errors 
with the O9.5V case. In the case of the latter distance, the evidence for \hd~ to lie 
in front of the cluster, would be much weaker.

We conclude that \hd~ may lie up to a few hundred pc in front of the NGC 6530 cluster.
 However, we cannot exclude that the star is located inside or at the front of the cluster. 
The peculiar radial velocity (corrected for Galactic rotation and solar motion using a local standard of rest 
of $(u,v,w)_{\odot} = (10.4, 11.6, 6.1)$ km/s (Tetzlaff et al. 2011)) of HD 164816 is $v_{r,pec}=-80{.}7^{+5.0}_{-4.4}\,
\mathrm{km/s}$ whereas for NGC 6530 it is $v_{r,pec}=-7{.}9^{+3.5}_{-9.1}\,
\mathrm{km/s}$ (Kharchenko et al. 2005). 
Hence, HD 164816 appears to move towards us relative to the NGC 6530 cluster. 
The difference between the two velocities is $72.8 \pm 7.9$ km/s.
 While the radial velocity difference indicates that the star is now moving towards us relativ to the cluster,
it can still be located inside or at the front edge of the cluster. If the system includes a neutron star born
in a supernova, this supernova should have given the system a kick, which may have been the cause for the discrepant
radial velocity. Given that the most massive and earliest star in this cluster (9 Sgr) has a spectral type of O4,
the progenitor of the neutron star in HD 164816 has to have had an even earlier spectral type and, hence, had
a life-time of below 3 Myr. Given that this is comparable to the cluster age (2.6 Myr), the presumable
supernova should have happened very recently and the system HD 164816 therefore cannot be located much foreground
to the cluster.
The proper motion of HD 164816 actually agrees with the typical proper motion of the NGC 6530 cluster.

\subsection{X-ray longterm variability}

As we have discovered a $\sim$3.81 d orbital period we searched the available X-ray data
 sets mentioned in \S2 for any longterm variability; in particular those are three \emph{Chandra} 
and one \emph{XMM-Newton} set.\\
For this purpose we first derived the absorbed fluxes for \emph{XMM-Newton} and \emph{Chandra} observations
(see \S2.1 and \S2.2) and plotted them vs the starting time of each observation taken from table \ref{X_obs}. 
As there is only very limited statistics in each of the three \emph{Chandra} exposures, we merge them and use the 
starting time of the middle observation. As a result  
one can see that the fluxes are consistent within the error bars.
We have as well investigated the longterm behavior of the individual \emph{Chandra} count rates, which cannot 
be compared to the \emph{XMM-Newton} count rate as their instrumental responses are different. We cannot find any 
variation in those count rates as well.  
Thus the merging of the datasets is justified. We hence cannot detect the $\sim$3.81 d orbital period 
in the available X-ray observations.

\section{Radio and $\gamma$-ray observations}
In order to get a complete view of the multiple system \hd~we have searched as well for any
significant detection in the available radio data catalogs. We could neither detect it in
the National Radio Astronomy Observatory (NRAO) Very Large Array (VLA) Sky Survey (NVSS) nor
by the GTEE 35 MHz Radio survey carried out by the low frequency T-array near Gauribidanur, India.
Some feature could be detected in the H~I All-Sky survey, the CO galactic plane survey and the
4850 MHz Survey carried out by the Parks and Green Bank Radio observatories. Due to
the rather large pixel scales of around one degree, 0.1 degree and 0.02 degree respectively
we consider the detected feature unlikely to be associated with \hd. In addition \hd~with
a galactic latitude of $\sim$ -1\degr~is lying quite close to the galactic plane and
the detected radio emission is hence more likely to come from diffuse emission of interstellar
gas.\\
With the \emph{FERMI} $\gamma$-ray telescope now in orbit we took as well data obtained by its
\emph{Large Area Telescope (LAT)} at the position of \hd~into consideration. As the whole sky
is monitored every six hours for a bit more than two years now we analyzed data taken over this
whole range (from 2008 Aug 04 to 2010 Aug 09) in order to get the highest possible statistics.
Events in the energy range from 600 MeV to 300 GeV have been chosen and 
have been filtered  
for galactic and extra galactic diffuse $\gamma$-ray emission. A
likelihood analysis has been carried out afterwards which serves as a source detection since the
photon statistics at the position of \hd~are still very low. For this purpose spectral models
for all sources in a ROI of ten degrees radius around \hd~have been accounted for contribution and
were modeled simultaneously. 
A super exponential cutoff powerlaw has been assumed for
\hd~which is typical for $\gamma$-ray emitting neutron stars (c.f. Trepl et al. 2010). Carrying out the
likelihood analysis yielded no significant detection of the source. 

\section{Discussion}
The X-ray source associated with \hd~shows X-ray pulsations at $\sim$ 9.78 s and
in addition a soft X-ray excess with a blackbody temperature of $\sim$ 49 eV both 
consistent with a compact companion (NS) in the system. Further optical
observations confirm that \hd~is a spectroscopic binary consisting of two O9V or O9.5V stars
that are in a tight orbit of $\sim$ 3.82 days.  
\\
Accreting neutron stars and generally neutron stars in high-mass and supergiant X-ray binaries have in common that 
they emit at higher energies than 2 keV,
but in the EPIC PN observation we did not detect any signal at above
2 keV, so that we can exclude an accreting pulsar as the source for the X-ray emission.\\ 
All the aforementioned scenarios have in common that the NS is
emitting non-thermal radiation that can be fitted by a powerlaw or Bremsstrahlung-model
and is thought to occur from the accretion process (either Roche-lobe overflow or wind accretion).
Constant mass accretion can in fact be ruled out in the case of \hd~as the detected soft X-ray emission excess
from \xmmssc~can be described by a pure blackbody model which means we observe
thermal emission from the NS surface itself. In addition, evolutionary considerations 
based on standard formulae (Lipunov 1992), demonstrate that the stage of accretion is
improbable taking into account the spin period of the NS and the parameters of the binary. 
Without better knowledge about the separation of the NS from the O-stars it is impossible to make 
clearer statements, but at a few Myr the NS is either in the ejector, or in the propeller stage.\\
The orbital separation between the compact companion and the binary system must
be at least around 100 R$_{\odot}$ as empirical studies show that a tighter system is
dynamically unstable unless the separation of the compact object from the
binary center of masses is about three to five times larger than the semi-major axis of the
binary system for an eccentricity of $\sim$ 0.2 (Moriwaki \& Nakagawa 2002) i.e. $\geq$ 51 to 85 R$_{\odot}$ here.
Therefore wind accretion is negligible as the orbital separation is large.\\
The other possible nature of the compact companion to \hd~might be a White Dwarf (WD) as the detected
temperature of $\sim$ 50 eV (5.80$\times$10$^{5}$ K) would place the object
in the somewhat overlapping region of WDs as Super-Soft X-ray sources (SSS) in
Be/X-ray binaries (Kahabka et al. 2006) and thermally emitting NS.\\
However as the typical age of a WD should be $\gtrsim$ 1 Gyr the accompanying O-stars
with masses of around 18 M$_{\odot}$ and 20 M$_{\odot}$ respectively would have already evolved and exploded.\\
The compact companion lying just by chance in the line of sight of \hd~cannot be ruled out yet as
the available \emph{Chandra} observations are only spread
by a few days (see Table \ref{X_obs}). This is too short to significantly detect any separation between the
centroid of the soft and hard component of the X-ray spectrum.\\
If this detection is a chance projection, then the most probable candidates are  
radio pulsars.
The probability for chance alignment of a neutron star within the
point spread function (PSF) of the X-ray source can be estimated as follows:
Neutron stars can be detected as either X-ray, $\gamma$-ray or radio pulsars
if they lie above the so-called dead-line in the P - $\dot{P}$ diagram; 
they reach this dead-line at an age
of roughly 10 Myr (despite according to general cooling curves a NS may cool down to the temperature of $\sim$50 eV
at the age of $\sim$5$\times$10$^{5}$yr); if there are $5 \times 10^{8}$
neutron stars in the Galaxy (at ages up to 12 Gyr), then there are
roughly 3$\times$10$^5$ neutron stars detectable as either X-ray or radio pulsars
($\sim 2000$ of them are known). Then, given the PSF of the Chandra source
(full width at half maximum $3.70 \pm 1.60$ arcsec) or the PSF of the
XMM source (full width at half maximum $4.19 \pm 1.78$ arcsec)
and the area of the whole sky, we obtain a probability for a
chance alignment, between \hd~and a NS of 5 to $7 \times 10^{-4}$. 
However, the direction of these \emph{XMM} and \emph{Chandra} pointings is not random, but
towards an OB cluster, where the probability for a NS is higher: 
According to Hohle et al. (2010), the area on the sky where almost all 
SNe are expected (inside OB associations), is $35~\%$ of the total sky.
The probability for chance alignment of a NS in the 
\emph{XMM} and \emph{Chandra} PSF is then 2 to $3 \times 10^{-4}$, still very low.
Even if we restrict the estimate to the densest OB clusters close to the Galactic
plane, the estimate would be less than one order of magnitude higher, so that the
probability of chance alignment would still be below $1~\%$. 
Given that this estimate is very low, we have a large probability
for the potential NS not to be a chance alignment, but
to be related to HD 164816. This suggests that a NS is related to \hd.\\
O stars are generally not known to have soft X-ray blackbody excesses. It has been shown by
Naz\'e 2009 that those stars are in general emitting spectra that can be fitted best with single or
multi component MEKAL models with temperatures starting at $\sim$0.2 keV.\\
 A scenario that involves colliding winds instead of a compact companion can be considered as well
as \hd~consists of a close binary. However such a model would require a Raymond-Smith thermal
plasma with a temperature of kT $\sim$ 0.6 keV (Oskinova 2005) or a hard non-thermal X-ray component
(De Becker et al. 2004). Taking a look at the MEKAL temperature of $\sim$ 0.2 keV in the case of \hd~we note
that it is almost a factor of three lower. In addition the spectrum is missing any component that can be
fitted by a non-thermal model as the spectral counts from 2 keV on are consistent with zero. 
Hence, the NS does not appear to be accreting.\\
As seen from \S 4 \hd~is not significant at radio wavelengths; this might have implications for 
the colliding wind scenario (De Becker et al. 2004), but that is beyond the scope of this paper.\\
Assuming a distance 864$\pm$136 pc we compute the X-ray luminosity to L$_{X}\sim$ 2.58$\times$10$^{31}$erg s$^{-1}$. 
Together 
with a bolometric luminosity for a O9.5V star from Hohle et al. (2010) this yields log(L$_{X}$/L$_{Bol}$) = -6.73 which is 
in agreement with the relation for O stars found by Bergh\"ofer et al. (1997) (c.f. Fig. 4 therein). In the case of 
colliding winds this ratio should however be higher than the given value (see e.g. De Becker et al. 2004). 
It seems thus rather unlikely to get the low energy excess from colliding wind interaction. 
A similar conclusion was reached by Rauw et al. (2002), 
where they state that regarding the L$_{X}$/L$_{Bol}$ ratio no evidence for an increased value is found.\\
Like Rauw et al. (2002) we find a higher than average interstellar absorption (see table \ref{spec_par}) which
can be attributed to circumstellar or nebula material. This should result in strong suppression of the spectrum in 
the energy range 0.2 - 0.5 keV. In contrast to this we find an excess in that range on top of the absorption. Hence 
an additional source outside the region of circumstellar material responsible for that is highly probable. This is consistent 
with the earlier conclusion that the separation of the probable NS has to be at least 100 R$_{\odot}$.\\
Emission line features in the X-ray spectrum as described in van der Meer et al. (2005) are not detected and as well not 
 expected as such a phenomenon is only common in HMXBs and SGXBs. Those cases can be excluded as they both involve 
accretion processes which are not seen in our case.\\
If there is a pulsating NS within the X-ray PSF and if this NS is not orbiting the O star of \hd, then it could be an 
isolated NS in the NGS 6530 cluster. A pulsation period of some 10s is not atypical for isolated young to intermediate-age 
NS - as e.g. the Magnificent Seven with pulsation periods of few to some 10s (see e.g. Haberl 2007).\\

\section{Conclusion}
We found peculiarities in the X-ray, optical, and kinematical
data of HD 164816:
\begin{itemize}
\item There is a soft excess in the X-ray spectrum of both XMM and Chandra.
\item The radius of the emitting (circular) area related to the soft excess would be $\sim$7 km.
\item There is an indication of 9.78 s pulsation in the XMM data
(not detectable with Chandra due to low timing resolution of the used ACIS-I detector ($\sim$3.24 s) and small count rate).
\item If HD 184816 includes a neutron star born in a supernova, this supernova should have given the system
a kick, which is consistent with the fact that the star HD 164816 has a significantly different radial velocity
than the cluster mean.
\end{itemize}

All those four observational indications would be consistent with a compact object like
a neutron star in the system orbiting the spectroscopic binary HD 164816
at a larger separation. A NS should have soft X-ray emission.
A neutron star of few Myr age isolated from the
other stars without accretion can have a period of few to 10 s (like
the Magnificent Seven neutron stars).
If there is a NS in the system, there should have been a
supernova in the system before. Such a supernova could introduce a kick
velocity to the neutron star born in the supernova. If the neutron
star remained bound to the close spectroscopic binary, they would
now all move away fast relative to their birth cluster.

We did not find any evidence for X-ray emission due to colliding winds.
An accreting neutron star would show hard X-ray emission, which was not
found. Hence, if there is a neutron star in the system it is effectively
isolated (and non-accreting) from the O-type star.

\section*{Acknowledgments}
LT, VVH and MMH would like to thank the German \emph{Deut\-sche For\-schungs\-ge\-mein\-schaft, DFG\/}
for financial support in project SFB TR 7 Gravitational Wave Astronomy.\\
TP acknowledges support from the EU in the FP6 MC ToK project MTKD-CT-2006-042514 and Slovak
Academy of Sciences grant VEGA 2/0094/11.\\
NT acknowledges Carl-Zeiss-Stiftung for a scholarship.\\
RC would like to thank the Nordrhein-Westf\"alische Akademie der Wissenschaften und der K\"unste
in the framework of the academy program by the Federal Republic of Germany and the state
Nordrhein-Westfalen for support.\\
All authors would like to thank the numerous observers from the Bochum team: L.-S. Buda, P.
Bugeno, M. D\"orr, H. Drass, V.H. Hoffmeister, I. Lingner and R. Watermann.\\
Finally, we thank an anonymous referee for many good suggestions.

\label{lastpage}
\end{document}